# Enhanced electrical resistance at the field-induced magnetic transitions in some non-stoichiometric and stoichiometic Tb-based ternary germanides

K. Mukherjee*, Kartik K. Iyer, and E. V. Sampathkumaran
*Tata Institute of Fundamental Research, Homi Bhabha Road, Colaba, Mumbai 400005, India*

**Abstract**

We present the magnetic and transport behavior of some Tb compounds viz: $TbIrGe_2$, $TbFe_{0.4}Ge_2$ and $TbCo_{0.4}Ge_2$. The stoichiometric germenide $TbIrGe_2$ exhibits at least two distinct magnetic transitions in a close temperature interval around 10K. The non-stochiometric compounds, $TbFe_{0.4}Ge_2$ and $TbCo_{0.4}Ge_2$, undergo magnetic ordering around 17 and 19K respectively. The magnetic state of these compounds appears to be antiferromagnetic-like. Qualitatively there is a correlation between the field response of the magnetization (*M*), magnetoresistance (*MR*) and entropy change (*ΔS*) curve in all the three compounds. That is, these Tb compounds exhibit a *positive MR* and *ΔS* beyond a magnetic field where *M* also shows a field-induced transition. On the basis of this correlation, we conclude that magnetic disorder/fluctuations beyond a critical field - "a phenomenon called inverse metamagnetism"- rather than metamagnetism, is induced in these compounds.

## I. Introduction

In the field of condensed matter physics, the studies of rare earth intermetallics and oxides have attracted considerable attention over last few decades, as they exhibit exotic physical phenomena like colossal magnetoresistance, giant magnetocaloric effect, multiferroic properties etc. The fundamental reason for the existence of such multi-functional properties in these compounds is the coupling of magnetic subsystem with other degrees of freedom resulting from field-induced metamagnetic transitions. Recently we have reported giant positive magnetoresistance, defined as $[\rho(H)-\rho(0)]/\rho(0)$ (where $\rho$ is electrical resistivity and $H$ is the magnetic field), in an alloy $Tb_5Si_3$ [1]. It was seen that beyond a critical magnetic field, magnetic fluctuations are presumably induced at least at one of the magnetic sites of this antiferromagnetic compound- a phenomenon called "inverse metamagnetism" [2], which is not so commonly known in intermetallics. Such fluctuations lead to a higher resistive state, leading to the observed giant positive *MR* [2, 3]. It was also shown that this *MR* behavior changes with Lu substitution in $Tb_{1-x}Lu_xSi_3$, possibly giving rise to an unusual electronic phase separation comprising of high-field high-resistive phase and low-field low-resistive phase [4] in the zero field, after a field-cycling, unlike that observed in manganites or other intermetallic compounds [5]. It is therefore of interest to explore whether such a MR behavior is unique only to this compound. In this article, we report this behavior in another family of compounds, namely Tb based ternary germinides, $RY_xZ_2$, where R is a rare-earth element, Y is a d-electron element and Z is a p-electron element (Z = Si, Ge etc)

Among these ternary Tb compounds, silicides crystallize in the stoichiometric form (x=1), while germanides form both in stoichiometric or non-stoichiometric structure (0<x<1), the later one arising due to defects in d-electron metal sub-lattice [6]. The stoichiometric compounds crystallize in orthorhombic structure of YIrGe-type (Immm space group), in which the rare-earth ions occupy two non-equivalent lattice sites [7] with local symmetries *4i* and *4h*. The non-stoichiometric compounds crystallize in orthorhombic structure of $CeNiSi_2$-type (Cmcm space group) [8] and the local symmetry of Tb is *4c*. Thus it is seen that even though all the compounds form in the orthorhombic structure, there is a difference in the space group for the stoichiometric and non-stoichiomtric compounds. The non-stoichiometric compounds are more compact (that is, unit cells are of reduced volume) as compared to the stoichiometric compound. Among the germanides, Baran et al., [8] have reported the magnetic properties of $RCo_xGe_2$ (R = Gd-Er), and it was found that all but Tb and Dy show antiferromagnetic (AFM) ordering at low temperatures. Neutron diffraction measurements carried out on $TbNi_{0.4}Ge_2$ and $TbCu_{0.4}Ge_2$ showed that the rare earth orders antiferromagnetically below $T_N$ = 16 and 39 K respectively [9]. Some investigations were also carried out on polycrystalline $TbFe_{0.4}Ge_2$ [10]. For the stoichiometric structure, magnetic properties of the ternary compounds $RPtGe_2$ and $RIrGe_2$ (R=Gd-Er) have been reported [7, 11].

Here we present the results of detailed high-field magnetization (*M*), magnetic hysteresis, heat capacity (*C*) and magnetoresistance (*MR*) studies of $TbIrGe_2$ and non-stoichiometric $TbFe_{0.4}Ge_2$ and $TbCo_{0.4}Ge_2$. Since there are no literature reports on the transport behavior of these compounds, we have performed those studies as well. It is seen that the compound, $TbIrGe_2$, with two magnetic transitions at low temperatures, is characterized by a huge positive *MR* (about 50%) at the magnetic field at which there are magnetization jumps. A clear evidence of magnetic ordering is observed in $TbFe_{0.4}Ge_2$, unlike in Ref [10]. However, a weak negative

*MR* (<1%) is observed in this compound for an initial application of magnetic field, with *MR* decreasing at lower fields and then increasing at higher fields. The compound TbCo$_{0.4}$Ge$_2$ which *is being studied for the first time* also shows magnetic ordering at around 19K. In the magnetically ordered state at low temperatures, the *MR* is initially negative and crosses to a positive value beyond a certain field around which there appears to be a broadened transition in magnetization. The *MR* curve is symmetric for a reversal in field direction. The nature of the *MR* curve shown by this compound is quite intriguing. In short, a common feature in all these compounds is that there is a positive *MR* and entropy change, developing at a field at which the magnetization also shows a field-induced transition, mimicking the behavior of Tb$_5$Si$_3$.

## II. Experimental details

The compounds, TbIrGe$_2$, TbFe$_{0.4}$Ge$_2$ and TbCo$_{0.4}$Ge$_2$, were prepared by arc melting stoichiometric amounts of respective high purity elements in an atmosphere of argon. The ingots of the first compound were annealed at 560$^0$C for a week, while the latter two samples were annealed at 800$^0$C for the same duration. X-ray diffraction studies on the samples indicate that all the samples are crystallographically single phase (Fig 1) and the pattern collected is analyzed by the Rietveld profile refinement [12]. The *dc* magnetization (in the range 1.8-300 K) for all specimens was carried out with the help of a commercial superconducting quantum interference device (Quantum Design). *M(H)* measurements at selected temperatures (*T*) for all specimens were carried out by employing a commercial vibrating sample magnetometer (Oxford Instruments). We have performed heat capacity measurements employing a physical property measurements (PPMS) system by Quantum Design. The electrical resistivity ($\rho$) measurements by standard four probe method, in absence/presence of magnetic fields (upto 100 kOe, *T*=1.8–300 K) were performed with the same PPMS. A conducting silver paint was used for making electrical contacts of the leads with the samples.

The lattice constants of TbIrGe$_2$, found from Rietveld analysis are a=4.2714(1) Å, b=15.9850(4) Å and c=8.8333(2) Å. The lattice constants of TbFe$_{0.4}$Ge$_2$ and TbCo$_{0.4}$Ge$_2$ are a=4.1285(1) Å, b=15.8936(5) Å, c=4.0109(1) Å and a=4.1143(1) Å, b=16.0144(4) Å, c=4.0037(2) Å respectively. The "goodness of fit" of Rietveld refinement for all the compounds was around 1.3.

## III. Results and discussions:

### TbIrGe$_2$

Figure 2 shows the temperature dependence of magnetization of TbIrGe$_2$ measured in a field of 100 Oe. The plot reveals two closely spaced magnetic transitions, one around 12K and the other at 9K, which is more apparent from *d$\chi$/dT* verses *T* plot, where $\chi$ is the measured susceptibility (inset of Fig. 2). These features are also prevalent at a higher field of 5 kOe. There is no bifurcation of the zero field cooled (ZFC) and field cooled (FC) magnetization curves. The drop observed below 10 K is a typical signature of long range antiferromagnetism. Also no thermal hysteresis could be detected in the cooling and warming cycle, thereby ruling out any first order magnetic phase transition. The magnetic susceptibility of this compound obeys Curie-Weiss (*CW*) law above 50 K with the value of paramagnetic Curie temperature ($\theta_p$) and effective moment ($\mu_{eff}$) being nearly equal to -11 K and 9.8$\mu_B$ respectively. The value of $\mu_{eff}$ is close to that of Tb$^{3+}$ free ion (9.72 $\mu_B$) and the sign of $\theta_p$ is in accordance with the dominance of AFM interaction in agreement with a previous report [7]. These claims are further substantiated from the temperature dependence of *C* which shows distinct sharp peaks at around 12 K and 9 K; further, *C(T)* show an additional transition at 10 K (Fig. 3). The observation of sharp peaks is an indication of the presence of long range magnetic ordering. Viewed together with the magnetization data, it can be stated that this compound has at least two different magnetic ordering temperatures. The third peak at the intermediate temperature at 10 K is observed only in *C(T)* data indicating a subtle nature of magnetic ordering. However, only one peak is observed at 9.5 K (suppressing 9 and 10 K peaks) in a field of 50 kOe, indicating that the peaks at 9 and 10 K move in opposite directions with the application of magnetic field, as though the 9 K transition contains a ferromagnetic component and the 10 K transition is actually antiferromagnetic-like. The 12 K peak is found to be suppressed. At this juncture it is worth recalling [13] that a theoretical model was developed for the Gd systems to distinguish between commensurate and incommensurate magnetic structures and that multiple jumps in *C(T)* could be the result of these magnetic transitions. Qualitatively speaking, such ideas could be applicable to the present case as well, as the net change of C at these transitions appears small. A knowledge of crystal-field scheme is desirable to draw a more concrete conclusion in this regard.

From the above measurements, it appears that, for this compound there are three regions viz: a) above 12 K, b) between 12-9 K and c) an ordered region below 9 K. To further investigate these regions, isothermal magnetization curves were taken in these three different regions. The *M(H)* curves at 1.8 K (Fig. 4a) show a linear behavior up to 14 kOe and then a sudden rise is observed. Further a change of slope (or a kink) is observed at 29 kOe and 41 kOe. Beyond 41 kOe, a curvature is observed as if the magnetic moments are trying to align in the direction of the field. The overall shape of *M(H)* curve is reminiscent of metamagnetic systems. The magnetic moment at 120 kOe is around 5.3 $\mu_B$/formula unit which is far from saturation presumably due to crystal-field effects and/or due to inverse-metamagnetism for which further arguments will be advanced later in this article. A much higher field is required to attain the full Tb moment. A weak hysteresis is observed only below 41 kOe. The kinks are also observed at 5 and 8K also, however at a relatively lower field (Fig. 4b-c). Above the lower ordering temperature (9K), in the temperature interval of 9.5-13 K, smoothly varying *M(H)* curves are observed (Fig 4 d-f) and it is interesting that no difference is observed in the *M(H)* curves in this temperature range.

To have a better idea about the different regions of the compound, we now focus on the transport behavior. Figure 5(a) shows electrical resistivity as a function of temperature. In zero field, $\rho$ gradually increases with increasing temperature beyond 13 K. At a low temperature (~10 K), there is a sudden change in slope and $\rho$ tends to saturate well below 8 K. It is to be noted that temperature range of change of slope is the same as that in $\chi(T)$ curve. The signature of multiple transitions is however smeared in the resistivity curve. The slope change is not suppressed in a field of 50 kOe and it is seen that the resistivity is enhanced with applied field below 10 K, while it is suppressed above this temperature range. This observation is also reflected in the *MR(H)* behavior. At 1.8 K, the variation of *MR* up to 12 kOe is insignificant, beyond which *MR* rises sharply up to 50 kOe and reaches a value of about 50% near 50 kOe. Above this field, *MR* becomes nearly independent of field. Such a jump of *MR* above 12 kOe can be correlated to field-induced anomalies in *M(H)*. A similar variation of *MR* is observed at 9 K as well, but it is much weaker for 11 K with a crossover of sign of *MR* around 40 kOe. At 15 K, the sign of *MR* is negative, typical of a paramagnet. Interestingly, a striking similarity is observed between the *MR* and magnetocaloric effect curve, measured in terms of isothermal entropy change ($\Delta S$) (obtained by employing Maxwell's equations) [14], of this compound (Fig. 5c). Such a similarity has also been observed in other intermetallic compounds [15]. Here, $\Delta S$ show a positive value up to 9 K, as though magnetic fluctuations are introduced by the application of magnetic field beyond 12 kOe. Qualitatively, this correlates well with the nature of $M(H)$ and $MR(H)$ curves. Further the $\Delta S$ curve for 11 K shows a crossover in sign at nearly the same field range at which $MR$ reveals a sign change. At 15 K, $\Delta S$ curve behavior is the same as that observed in the $MR$ curve. Such a sign change in the $\Delta S$ curve has also been observed in other Tb compounds as well [16, 17].

Hence from the above data, it can be inferred that this compound orders magnetically in zero magnetic field, presumably in some kind of antiferromagnet below 9 K. The field-induced magnetic fluctuations, in other words 'inverse metamagnetism', are responsible for the observed $MR$ and $\Delta S$ features in this compound.

### $TbFe_{0.4}Ge_2$

Figure 6 shows the $\chi(T)$ for the compound $TbFe_{0.4}Ge_2$ measured in the presence of different magnetic fields. A clear signature of magnetic ordering is prevalent at all measuring fields, indicating magnetic ordering around 17 K. An inset of Fig. 6(a) shows the FC and ZFC curves of the sample in a field of 100 Oe. A bifurcation of these curves takes place below 17 K (a feature occasionally reported for some antiferromagnetic ordered systems arising from anisotropy and grain boundary effects) with the FC curve continuously rising with the decreasing temperature. These features provide definitive evidence for the presence of magnetic ordering in compound, in contrast to that reported in Ref [10]. Above the ordering temperature, Curie-Weiss law is obeyed with the $\theta_p \sim -38$ K and $\mu_{eff} \sim 9.9\mu_B$ (nearly equal to that of free $Tb^{3+}$ ion). The field response of magnetization at 1.8 K shown in Figure 6(b) illustrates a magnetic hysteresis both in positive and negative cycles of field as though there is a ferromagnetic component. A curvature in low fields (~10 kOe) is observed in the first cycle indicating the presence of a field induced effect. The magnetization monotonously increases with the increasing field with no sign of saturation as though the net magnetization is antiferromagnetic in character. With the increase in temperature, the hysteresis gradually diminishes, as expected (Fig 6c). From the nature of $M(H)$ curves, we infer that this compound could be a canted antiferromagnet. Further evidence for magnetic ordering in this compound comes from heat capacity measurements. As shown in the inset of Fig. 6a, a broad peak around 17 K is observed in the $C(T)$ curve which gets suppressed in a field of 50 kOe. The weak nature of the peak suggests a disorder broadened magnetic transition or a modulated magnetic structure [13]. The intensity of the jump in $C(T)$ at the magnetic transition (in zero field) is not so weak that it can be attributed to impurities, as otherwise x-ray diffraction pattern (the sensitivity of which is less than 2% for additional phases) should have revealed an additional phase. In absence of a suitable reference for the lattice contribution, it is difficult to infer the magnetic entropy change across the magnetic transition temperature.

Electrical resistivity of the compound exhibits a kink around 18 K (Fig. 7a), that is near the magnetic ordering temperature. Application of a magnetic field smears this feature. Figure 7b shows the field response of the $MR$ which illustrates that $MR$ at 1.8 K, and to some extent at 9 K, first decreases at low fields and then increases as the magnetic field increases. The initial decrease of $MR$ is attributable to the suppression of magnetic superzone gaps [18]. However, beyond 10 kOe, the positive contribution of $MR$ develops, which is attributable to the enhancement of spin fluctuations in the antiferromagnetic sub-lattice. At 9 K, the increase of positive *MR* is diminished, while at 15 K, *MR* continuously decreases with increasing field. It is possible that the contribution due to magnetic Brillouin-zone gap effect tends to dominate over the field-induced magnetic fluctuation effects with increasing temperature resulting in the observed *MR* behavior. The *ΔS* curve as a function of field is shown in Fig. 7c. *ΔS* at low temperatures, e.g. at 3 K, is initially weakly negative and, beyond 10 kOe, it changes its sign, similar to *MR* curve at low temperatures. The curve for all other temperatures in the magnetically ordered state are in the positive quadrant with the value decreasing with increasing temperature for a given value of *H*, indicating a decrease in the field-induced spin fluctuation contribution. This establishes our argument in terms of field-induced spin-fluctuations to explain the positive *MR* behavior at high fields.

### *TbCo$_{0.4}$Ge$_2$*

Figure 8a shows the temperature response of $\chi$ (under ZFC cooled condition) of this compound in different fields. The data reveals magnetic ordering around 19 K with another additional hump-like feature around 8 K. The signatures of onset of magnetic ordering persist even at high fields; however, the lower temperature feature is suppressed. From an inset of Fig 8a it is seen that the bifurcation of the ZFC and FC curves starts around the magnetic ordering temperature. The isothermal magnetization behavior at different temperatures are shown in figure 8 b and c and the *M(H)* curves are similar to that of TbFe$_{0.4}$Ge$_2$. A curvature in low field is observed in the first upward cycle indicating the presence of a field-induced effect. In zero- and in-field, the temperature response of *C* is shown in the inset of figure 8a. A broad peak around 19 K is observed in the *C(T)* curve which gets suppressed in high fields. The nature of the curve is the same as that of TbFe$_{0.4}$Ge$_2$. There is no other feature at lower temperatures and these could imply that the low temperature feature around 8 K in $\chi(T)$ could be due to a metastable magnetic state or the entropy associated with this transition is very small.

The temperature dependence of *ρ* of the compound in zero and in field is shown in Fig 9. The curve shows two features, one around 20 K and other at 8 K (obtained from *dρ/dT* vs. *T* plot) and these temperatures coincide with those obtained from the magnetization data. Therefore, we think that there is another magnetic transition with a small change in entropy. With the application of a field of 50 kOe, resistivity gets enhanced at lower temperature, and above 13 K, *ρ* is suppressed. *MR* of the compound at different temperatures, shown in Fig 10 (a-f), reveals very interesting features. At 1.8 K, the sign of *MR* is negative for initial applications of field. Interestingly, an upturn is observed beyond 15 kOe, and *MR* shows a sign crossover near 32 kOe. Again there is a change of slope around 56 kOe. Unlike the other two Tb compounds, a significant hysteresis is observed in this case which points to a broadened first-order magnetic transition with the application of field. As the temperature is increased, say at 5 K, the nature of *MR* curve is similar, however without any hysteresis. This implies that, at low temperatures, beyond a critical field, the contribution from field-induced spin fluctuation is dominating over the effect due to possible magnetic gap. At a higher temperature, say around 9 K, the crossover to the positive quadrant diminishes significantly. At 11 K, *MR* is negative and the nature of the curve is sine wave-like, which indicates that there is a competition between positive and negative contributions of *MR*. The upturn beyond 15 kOe is significantly reduced at 15 K, and at 25 K the *MR* behavior is like that of any paramagnet, decreasing gradually with increasing field. Such a type of *MR* loop in a compound is rather scarce. *ΔS* of this compound as a function of field is shown in Fig 11. The features in the curve are similar to that observed for TbFe$_{0.4}$Ge$_2$; establishing that the same arguments are applicable to both the cases.

## IV. Summary

We have investigated the properties of some non-stoichiometric and stoichiometric Tb based germanides. The stoichiometric compound TbIrGe$_2$ reveals two closely-space magnetic transitions. The low temperature magnetic phase of this compound shows a positive *MR* with a correlation with the *M(H)* and *ΔS(H)* curves. This behavior is ascribed to the field-induced spin-disorder (fluctuation) in the system. The non-stoichiometric compounds TbFe$_{0.4}$Ge$_2$ and TbCo$_{0.4}$Ge$_2$ compounds also order antiferromagnetically around 17 K and 19 K respectively. *MR* behavior of these compounds highlights the competition between negative (possibly due to magnetic-gap effects) and positive (due to field-induced spin fluctuation effect) contribution. The main point of emphasis is that in all these germanides, there is a positive *MR* developing at a field at which the magnetization also shows a field induced transition. These results reveal that Tb$_5$Si$_3$ may not be unique in its *MR* anomaly and that there is a need to consider the concept of 'inverse metamagnetism' in general in antiferromagnetic systems.

*E-mail address: kmukherjee@.tifr.res.in

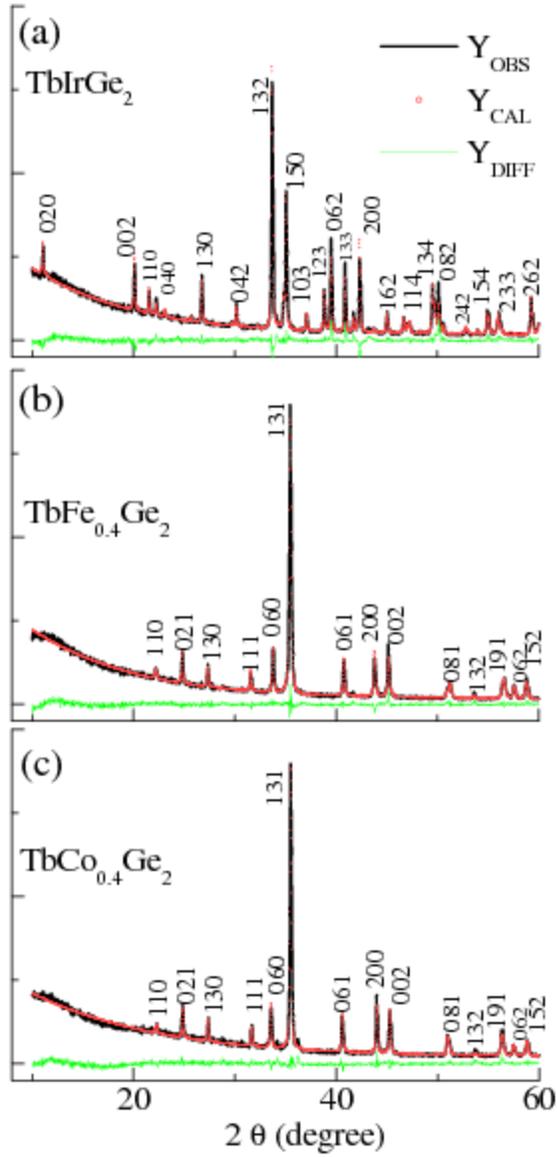

Figure 1:
(color online) X-ray diffraction patterns (Cu K$_\alpha$) for a) TbIrGe$_2$, b) TbFe$_{0.4}$Ge$_2$, and c) TbCo$_{0.4}$Ge$_2$. The observed [Y$_{OBS}$] (from experiment), calculated [Y$_{CAL}$] (from Rietveld analysis) and the difference curve [Y$_{DIFF}$] (between observed and calculated) also shown.

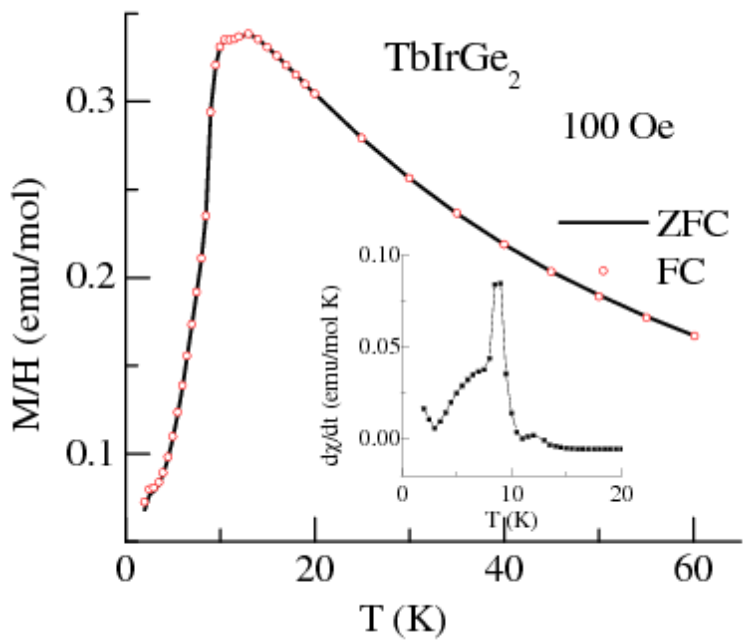

Figure 2:
(color online) Temperature (*T*) dependence (below 60 K) of magnetization (*M*) divided by magnetic field (*H*) for TbIrGe$_2$ for zero field cooled (ZFC) and field cooled (FC) case. Inset: Temperature response of d$\chi$/dT for the same at 100 Oe .

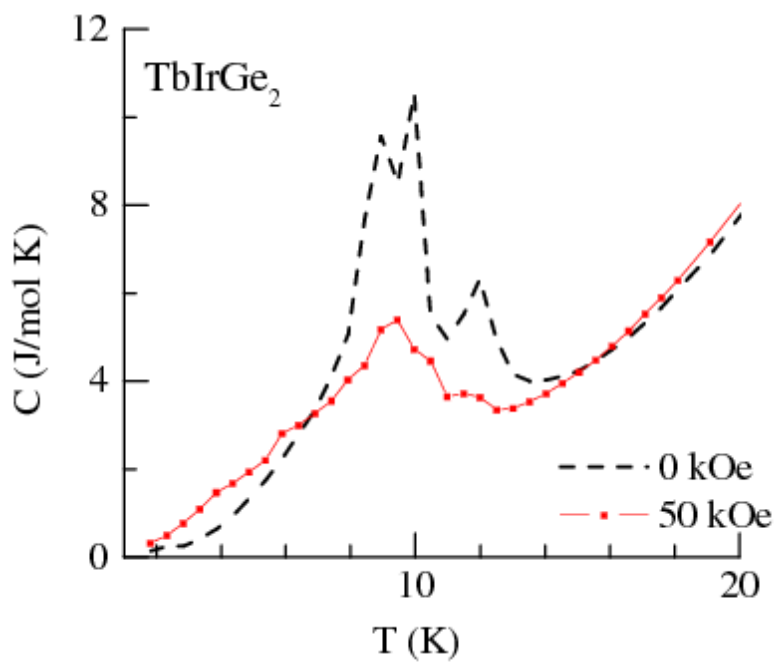

Figure 3:
(color online) Heat capacity of TbIrGe$_2$ as a function of temperature in zero field and in 50 kOe field.

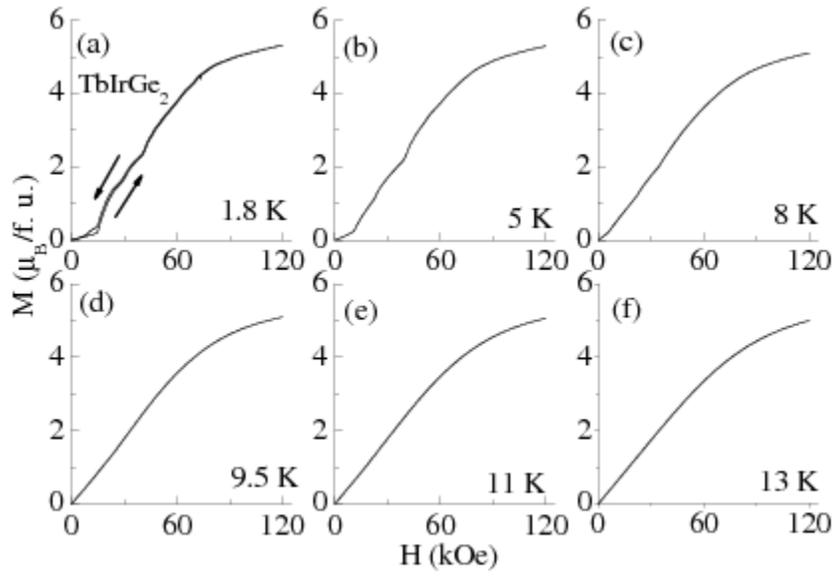

Figure 4:
(color online) Isothermal magnetization behavior of TbIrGe$_2$ at different temperatures.

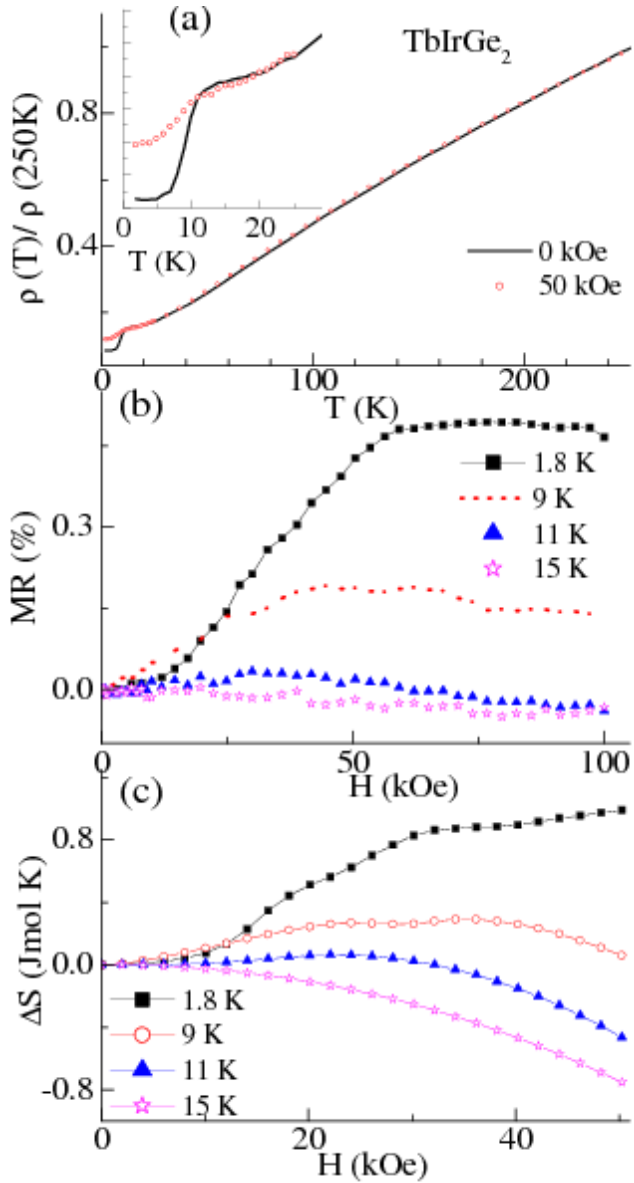

Figure 5:
(color online) (a) Electrical resistivity as a function of temperature in zero field and 50 kOe field. Inset shows the same upto 30K. (b) Magnetoresistance behavior of the sample as a function of field at different temperatures. (c) Isothermal entropy change as a function of field at the corresponding temperature is shown. The lines through the data points are guides to the eyes.

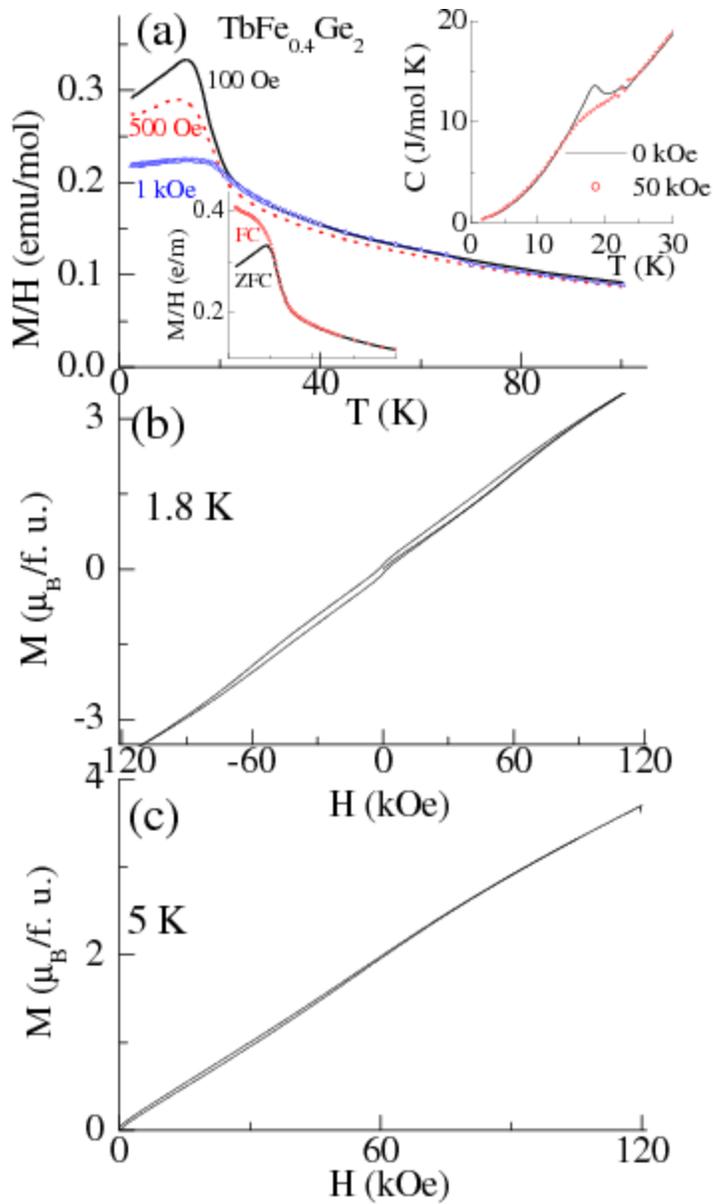

Figure 6:
(color online) (a) Temperature ($T$) dependence (below 90 K) of magnetization ($M$) divided by magnetic field ($H$) for TbFe$_{0.4}$Ge$_2$ at different fields. Left bottom inset: Zero field cooled and field- cooled magnetization of the compound measured in 100 Oe. Right top inset: Heat capacity of the sample as a function of temperature in zero field and 50 kOe field. (b) and (c) Isothermal magnetization of the compound at 1.8 and 5 K respectively.

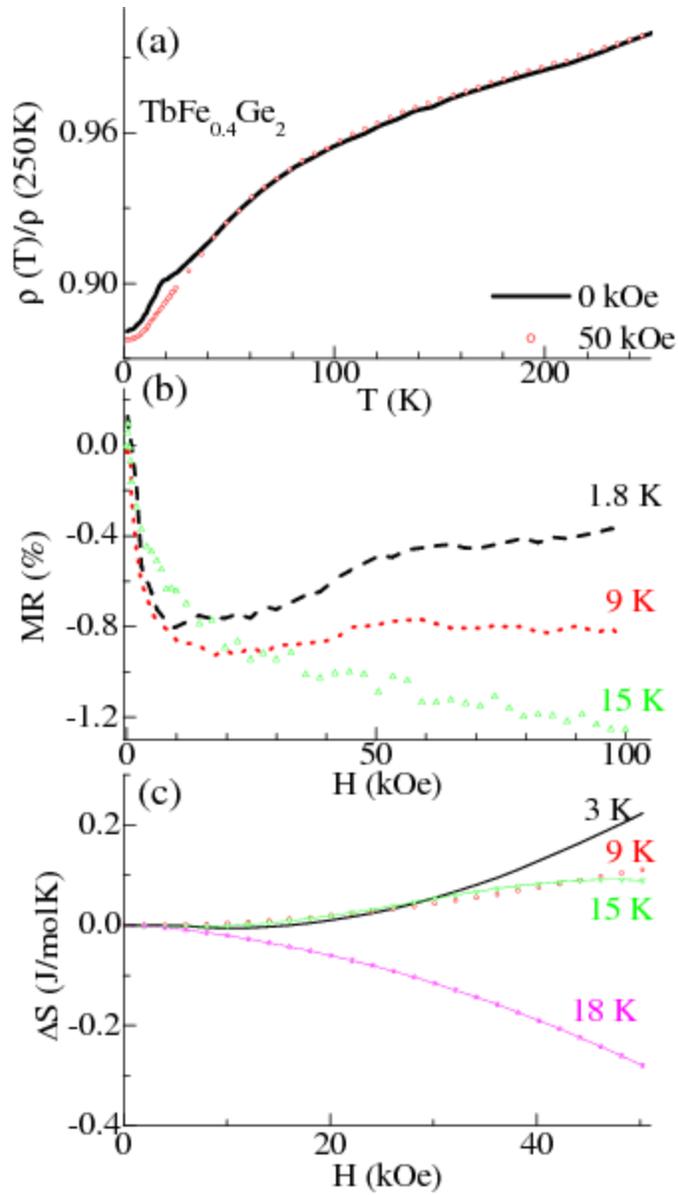

Figure 7:
(color online) (a) Electrical resistivity as a function of temperature in zero field and in 50 kOe field. (b) Magnetoresistance behavior of the compound as a function of field at different temperature. (c) Isothermal entropy change as a function of field at different temperature.

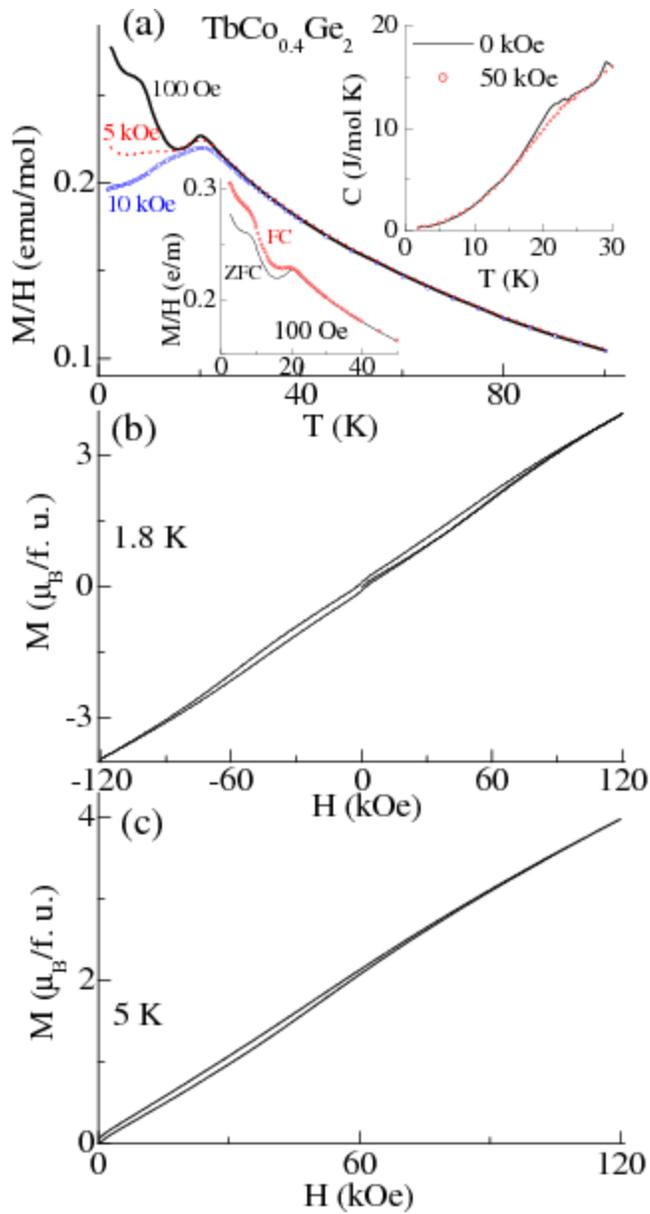

Figure 8:
(color online) (a) Temperature ($T$) dependence (below 90 K) of magnetization ($M$) divided by magnetic field ($H$) for TbCo$_{0.4}$Ge$_2$ measured in different fields. Left bottom inset: Zero field cooled and field-cooled magnetization behavior in 100 Oe. Right top inset: Heat capacity of the sample as a function of temperature in zero field and in 50 kOe field. (b) & (c) Isothermal magnetization of the compound at 1.8 K and 5 K respectively.
15

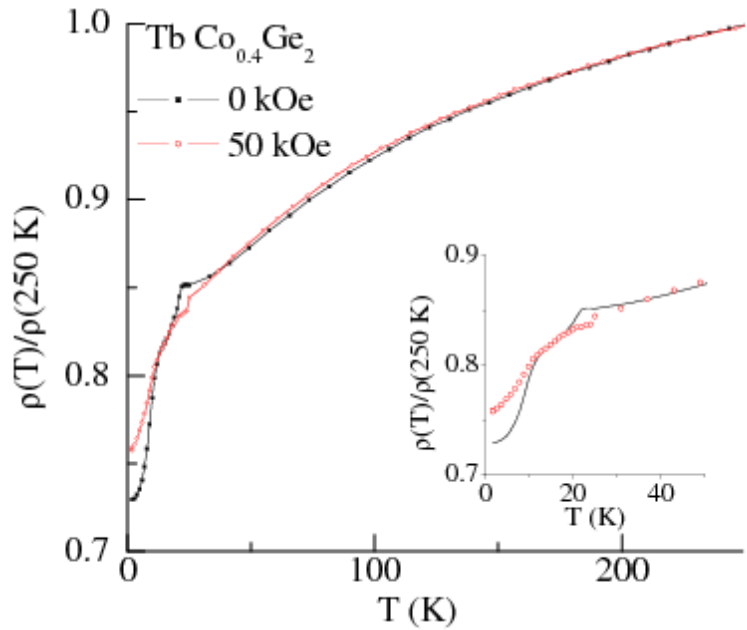

Figure 9:
(color online) Electrical resistivity as a function of temperature in zero field and in 50 k Oe field for TbCo$_{0.4}$Ge$_2$. Inset shows the same upto 40 K.

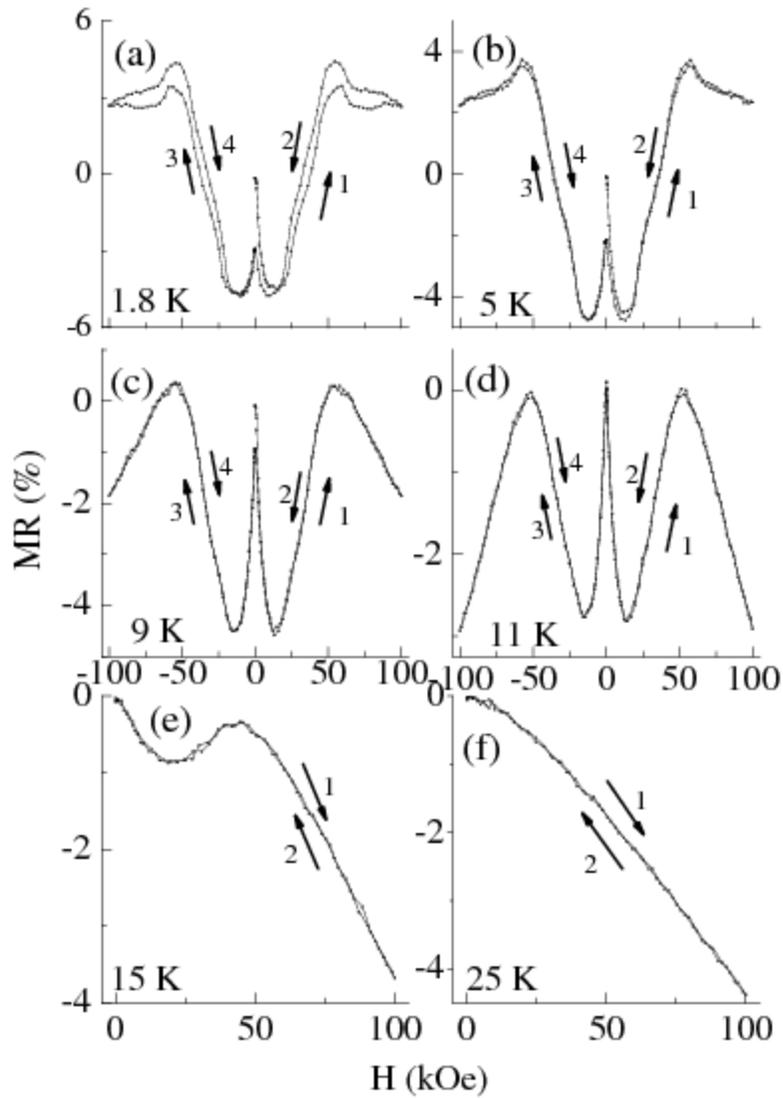

Figure 10:
Magnetoresistance as a function of field for TbCo$_{0.4}$Ge$_2$ at different temperatures. The arrows and numericals are placed on the curves as a guide to show the direction in which the magnetic field has been changed.

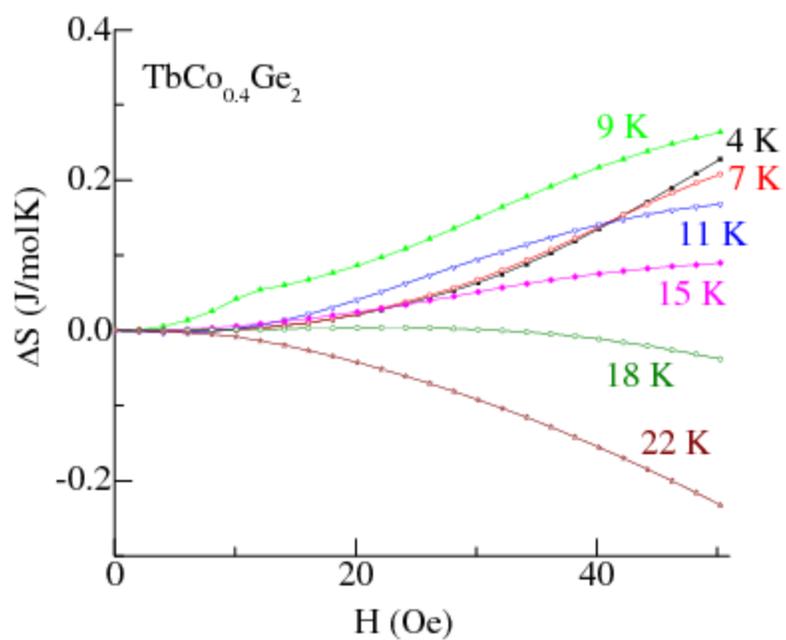

Figure 11
Isothermal entropy change for TbCo$_{0.4}$Ge$_2$ as a function of field at different temperatures.